\documentclass{mn2e}

\pdfoutput=1

\usepackage{fixltx2e}
\usepackage{mathptmx}
\usepackage[pdftex]{graphicx}

\def\cm{{\rm\thinspace cm}}

\def\erg{{\rm\thinspace erg}}

\def\K{{\rm\thinspace K}}
\def\keV{{\rm\thinspace keV}}
\def\km{{\rm\thinspace km}}
\def\kpc{{\rm\thinspace kpc}}

\def\Mpc{{\rm\thinspace Mpc}}

\def\s{{\rm\thinspace s}}
\def\yr{{\rm\thinspace yr}}

\def\ergps{\hbox{$\erg\s^{-1}\,$}}

\def\kmps{\hbox{$\km\s^{-1}\,$}}
\def\kmpspmp{\hbox{$\km\s^{-1}\Mpc{-1}\,$}}

\def\pcm{\hbox{$\cm^{-3}\,$}}

\begin{document}

\title[A wider view of the Perseus Cluster]{A wide Chandra view of
  the core of the Perseus cluster} \author[Fabian et al]
{\parbox[]{6.5in}{{A.C. Fabian$^1\thanks{E-mail: acf@ast.cam.ac.uk}$,
      J.S. Sanders$^1$,  S.W.~Allen$^2$, R.E.A.~Canning$^1$, E.~Churazov$^3$,
      C.S.~Crawford$^1$, W.~Forman$^4$, J.~GaBany$^5$,
      J.~Hlavacek-Larrondo$^1$, R.M.~Johnstone$^1$,
      H.R.~Russell$^6$, C.S.~Reynolds$^7$, P.~Salom\'e$^8$, G.B.~Taylor$^9$ and
      A.J.~Young$^{10}$
    }\\
    \footnotesize
    $^1$ Institute of Astronomy, Madingley Road, Cambridge CB3 0HA\\
    $^2$ Kavli Institute for Particle Astrophysics and Cosmology,
    Department of Physics, Stanford University, 452 Lomita Mall,
    Stanford, CA 94305-4085, and SLAC National Accelerator Laboratory,
    2575 Sand Hill Road, Menlo Park, CA 94025, USA
    \\
    $^3$ Max-Planck-Institut for Astrophysics,
    Karl-Schwarzschild-str. 1,
    85741 Garching, Germany\\
    $^4$ Harvard-Smithsonian Center for Astrophysics, 60 Garden St.,
    Cambridge, MA 02138, USA \\
    $^5$ Blackbird 2 Observatory, Alder Springs, CA93602, USA\\
    $^6$ Dept. of Physics and Astronomy, University of Waterloo, ON
    N2L 3G1, Canada\\
    $^7$ Dept. of Astronomy, University of Maryland, College Park,
    MD20742, USA\\
    $^8$ LERMA, Observatoire de Paris, UMR 8112 du CNRS, 75014 Paris, France\\
    $^9${University of New Mexico, Dept. of Physics and
      Astronomy, Albuquerque, NM 87131, USA; gbtaylor@unm.edu}\\
    $^{10}$ H. H. Wills Physics Laboratory, University of Bristol,
    Tyndall Avenue, Bristol BS8 3HG\\
  } }

\maketitle
  
\begin{abstract}
  We present new Chandra images of the X-ray emission from the core of
  the Perseus cluster of galaxies. The total observation time is now
  1.4~Ms. New depressions in X-ray surface brightness are discovered
  to the north of NGC\,1275, which we interpret as old rising
  bubbles. They imply that bubbles are long-lived and do not readily
  breakup when rising in the hot cluster atmosphere. The existence of
  a 300~kpc long NNW--SSW bubble axis means there cannot be
  significant transverse large scale flows exceeding
  $100\kmps$. Interesting spatial correlations are seen along that
  axis in early deep radio maps. A semi-circular cold front about
  100~kpc west of the nucleus is seen. It separates an inner disturbed
  region dominated by the activity of the active nucleus of NGC\,1275
  from the outer region where a subcluster merger dominates.
\end{abstract}


\section{Introduction}

New large-scale Chandra data on the core of the Perseus cluster,
A\,426, are presented here.  They follow our earlier 900~ks ACIS-S
image which covered a 8x8 arcmin region centred on the nucleus of the
dominant galaxy NGC\,1275 (Fabian et al 2000; Fabian et al 2006;
Sanders \& Fabian 2007). That image, with hints from neighbouring CCD
chips, indicated that bubbles, filaments and other fine-scale
structure extended over a wider field. The new data extend the area
imaged in detail to a region over 30~arcmin by 15~arcmin. XMM
observations of the Perseus cluster, of lower spatial resolution but
covering an even wider area than that discussed here, have been
presented and discussed by Churazov et al (2003).

\begin{figure}
  \centering
  \includegraphics[width=0.97\columnwidth,angle=0]{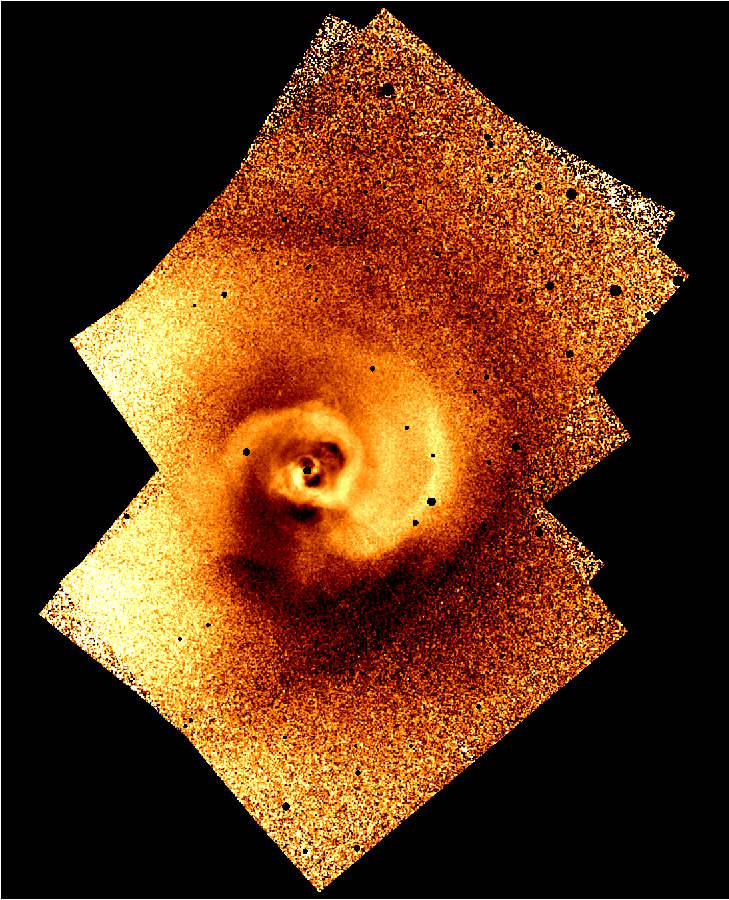}
  \caption{Mosaic of new images only, exposure corrected and adaptively
    smoothed.
The light streaks to the SW direction are due to the CCD readout of
the bright nucleus.  The
    separate ACIS-I pointings are evident, each has a side of 16~arcmin. }
\end{figure}

The Perseus cluster is the X-ray brightest cluster in the Sky (Forman
et al 1973), having peaked emission first resolved with the Copernicus
satellite (Fabian et al 1974). Einstein X-ray imaging showed a dip in
the X-ray emission to the NW of the nucleus of the central galaxy,
NGC\`1275, (Branduardi-Raymont et al 1981; Fabian et al 1981). ROSAT
HRI data (B\"ohringer et al 1993) showed that this dip was an outer
bubble in the hot X-ray emitting gas and resolved two inner bubbles
coincident with the FRI radio source, 3C84 (Pedlar et al
1990). Churazov et al (2000) showed that the bubbles are inflated by
the radio jets and rise buoyantly in the surrounding hot gas. The NW
bubble was formed a few $10^7\yr$ ago and is now moving
outward. Estimates of the likely energies and timescales indicated the
power in the bubbling process to be of the order of
$10^{45}\ergps$. Chandra images resolve the bubbles in detail (Fabian
et al 2000) and show subtle ripples centred on the inner bubbles which
are interpreted as weak shocks or sound waves produced by the bubbles
(Fabian et al 2003a). The shape of the NW bubble resembles the
cross-section of a rising spherical cap bubble in water on Earth
(Fabian et al 2003b), despite being over $10^{22}$ times
larger. Rising bubbles drag out cooler X-ray emitting gas (as in the
Virgo cluster, Churazov et al 2001) from the centre and cold gas (the
H$\alpha$ filaments of NGC\,1275: Lynds 1970; Conselice et al 2001;
Fabian et al 2003b; Fabian et al 2008).

In this paper we present evidence for further structures along the
bubbling axis which may be older, outer bubbles, or merged bubbles.

The cluster is assumed to have a redshift of 0.0183, corresponding to
an angular scale of 0.37 kpc arcsec$^{-1}$ ($H_0=70\kmpspmp$).   

\section {The new images}

The new ACIS-I images have been mosaicked together into a single image
0.5--7~keV shown in Fig.~1. The average at each radius has been subtracted for
display purposes. The
central structure is associated with the bubbles formed by the radio
jets from the active nucleus and is detailed in our earlier work.  The
separate ACIS-I chips are 16x16 arcmin and the whole image is 35.5
arcmin from N to S (788~kpc).

\begin{figure*}
  \centering
   \includegraphics[width=0.9\textwidth,angle=0]{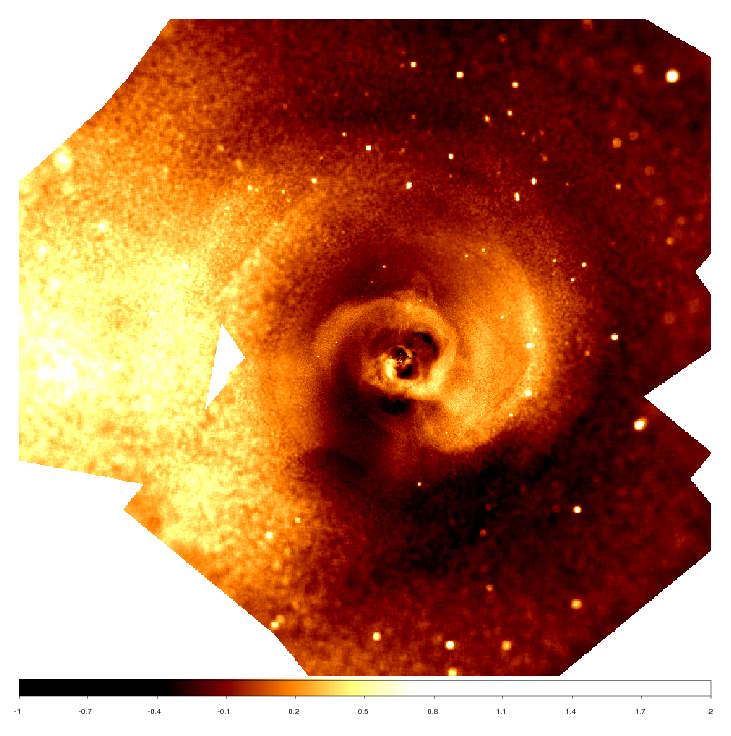}
   \caption{Final composite image of the 1.4~Ms exposure in the
     0.5--7~keV band, adaptively smoothed with a kernel with a fixed
     number of counts (225 in each of 3 energy bands, which have then
     been summed) and with the average at each radius subtracted. The
     colour shows fractional variation. The image is 25.6 arcmin from
     N to S. The optically second brightest galaxy in the cluster,
     NGC\,1272, lies along the bright rim 5 arcmin (111~kpc) W of the
     nucleus. }
\end{figure*}

We have added the earlier 900~ks ACIS-S images to make the final
composite, Fig.~2.  The nucleus of most member galaxies of the Perseus
cluster are detected (see Santra et al 2007). The point source in the
image to the WSW of the nucleus of NGC\,1275 (about halfway to the edge)
corresponds to the nucleus of the optically second-brightest cluster
galaxy, NGC\,1272 (see Fig.~3). It sits on a rim of X-ray bright
cluster emission.  This rim is seen as a step at 6--7 armin radius
from the nucleus of NGC\,1275 in Fig.~4 and 5, and is roughly circular
in shape with a centre displaced 1~arcmin to the NW of that
nucleus. We return to the large scale structure to the W in Section
6. 

Fig.~3 right was produced with the 0.5~m telescope at the Blackbird
Observatory, located in the south central Sacramento Mountains of New
Mexico USA. The image represents over 31~hr of cumulative exposure
time through broadband clear, red, green, blue and 6~nm narrow band
filters.  Each sub-exposure was reduced following standard procedures
for bias correction and flat fielding. Master dark and bias frames
were created by combining 10 dark sub-exposures each produced at the
same exposure length and camera temperature settings used for the
luminance and the filtered images. A master flat was produced by
combining 10 separate sky flat exposures for each filter. The clear,
red, green, blue and 6nm narrowband filtered exposures were separately
combined (using a median procedure) to produce individual master data
sets.  Each master data set was then projected into a gray scale
image, intensity scaled, assigned to its appropriate primary color
then digitally combined with the other data sets to produce a full
color picture.  The 6~nm narrowband data set was used to supplement the
intensity of the red broadband color channel; the clear data set was
used to intensify luminance information for the combined data sets.
Post processing methods that resulted with the final image followed
standard procedures that are also employed to create Hubble Heritage
images (Rector et al 2007).

\section{New Structures}
Overlays of the new composite X-ray image with the H$\alpha$ filaments
around NGC\,1275 and deep radio data at 49~cm (Sijbring 1993) are shown in
Figs. 7 and 8.  As noted by Fabian et al (2006) there is a looplike
X-ray structure at the end of the long Northern H$\alpha$
filament. This could be fallback of the outer parts of the cooler gas
dragged out to the N by earlier bubbles, when the optical filament 
was created. 

Further structures to the N are then expected and indeed
observed. The radio image shows an extension at the end of the optical
filament above which there is a spur leading to a dark patch in the
X-ray image, corresponding to a drop in flux, which we call the
Northern trough.  (Burns et al 1994 published a deep 330~MHz VLA radio 
image of Perseus which
shows similar outer structure, lending credence to the details being
real.) Just to the W of the radio spur lie two
elliptical-shaped dips in the X-ray image which we
identify as old outer bubbles. The outer of these was found originally
on an outer chip of the deepest ACIS-S observations (Sanders \& Fabian
2007). These structures are indicated in Fig.~9 and shown in profile in
Fig.~10.

\begin{figure*}
  \centering
  \includegraphics[width=0.96\textwidth,angle=0]{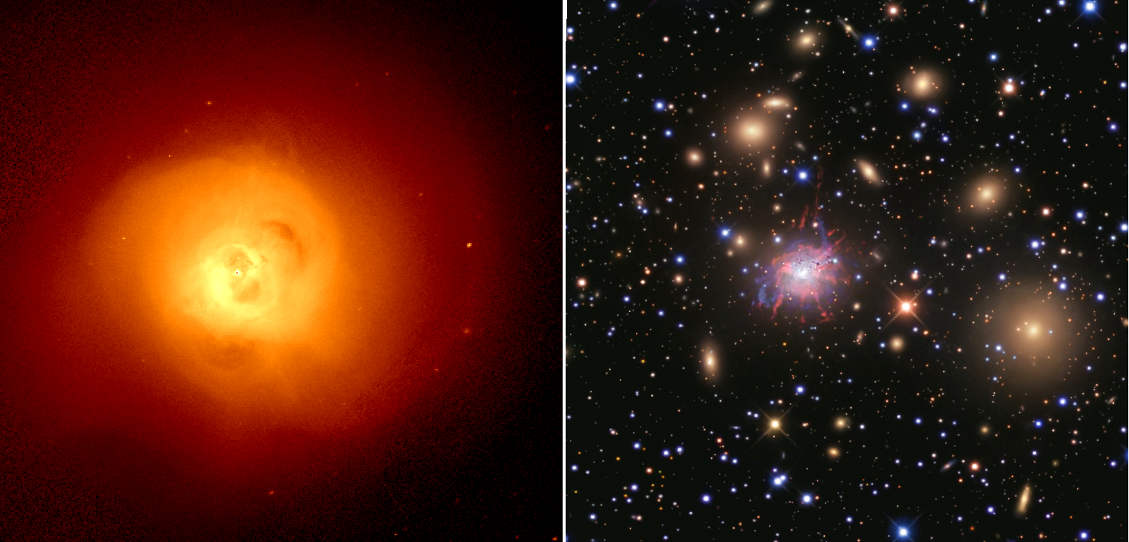}
  \caption{Matching X-ray and optical images of the core of the
    Perseus cluster  Left: Chandra composite 
(from Fig.~2, but without
    subtraction of the mean at each radius); Right:
    optical from Blackbird Observatory (see text for details). The
    images are 11.8 arcmin from N to S. NGC\,1272 is the bright
    elliptical galaxy 5 arcmin WSW of NGC\,1275.  }
\end{figure*}

A bay is seen to the SSE of NGC\,1275 (Fabian et al 2006), along the
bubble axis (Figs.~8 and 9). The radio contours indicate a coincident
minimum in radio flux.

\subsection{The Northern Trough and Bubbles}

The N trough (structure labelled A in Fig.~6) lies 9 arcmin (200~kpc)
N of the nucleus. We have studied it through profiles made by spectral
analysis along a sector to the N (Fig.~12). It is about 10 percent
deep in surface brightness and likely lies along, or close to, an
equipotential. It is also along the main jet axis to the N where
within the inner 40~kpc or so there is a pronounced optical H$\alpha$
filament (Figs.~3 and 7). If such a filament has been dragged out by
rising large bubbles (see Fabian et al 2003; Hatch et al 2006) then it
is plausible that the trough could be the remains of those
bubbles\footnote{The possibility that the jet has changed direction or
precessed is discussed by Dunn, Fabian \& Sanders (2006). The mean
axis over the past $\sim 5\times 10^8\yr$ remains along the NNW-SSW
direction.}.

\begin{figure}
  \centering
  \includegraphics[width=0.45\textwidth,angle=0]{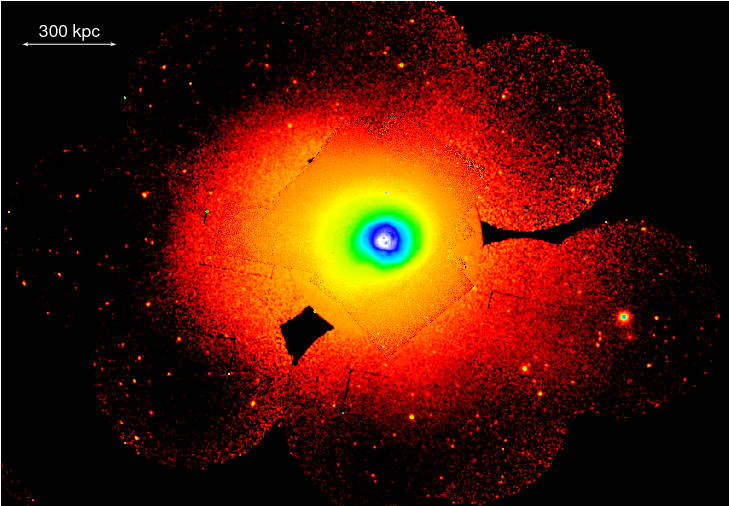}
  \caption{Joint Chandra and XMM image.}
\end{figure}

\begin{figure}
  \centering
  \includegraphics[width=0.45\textwidth,angle=0]{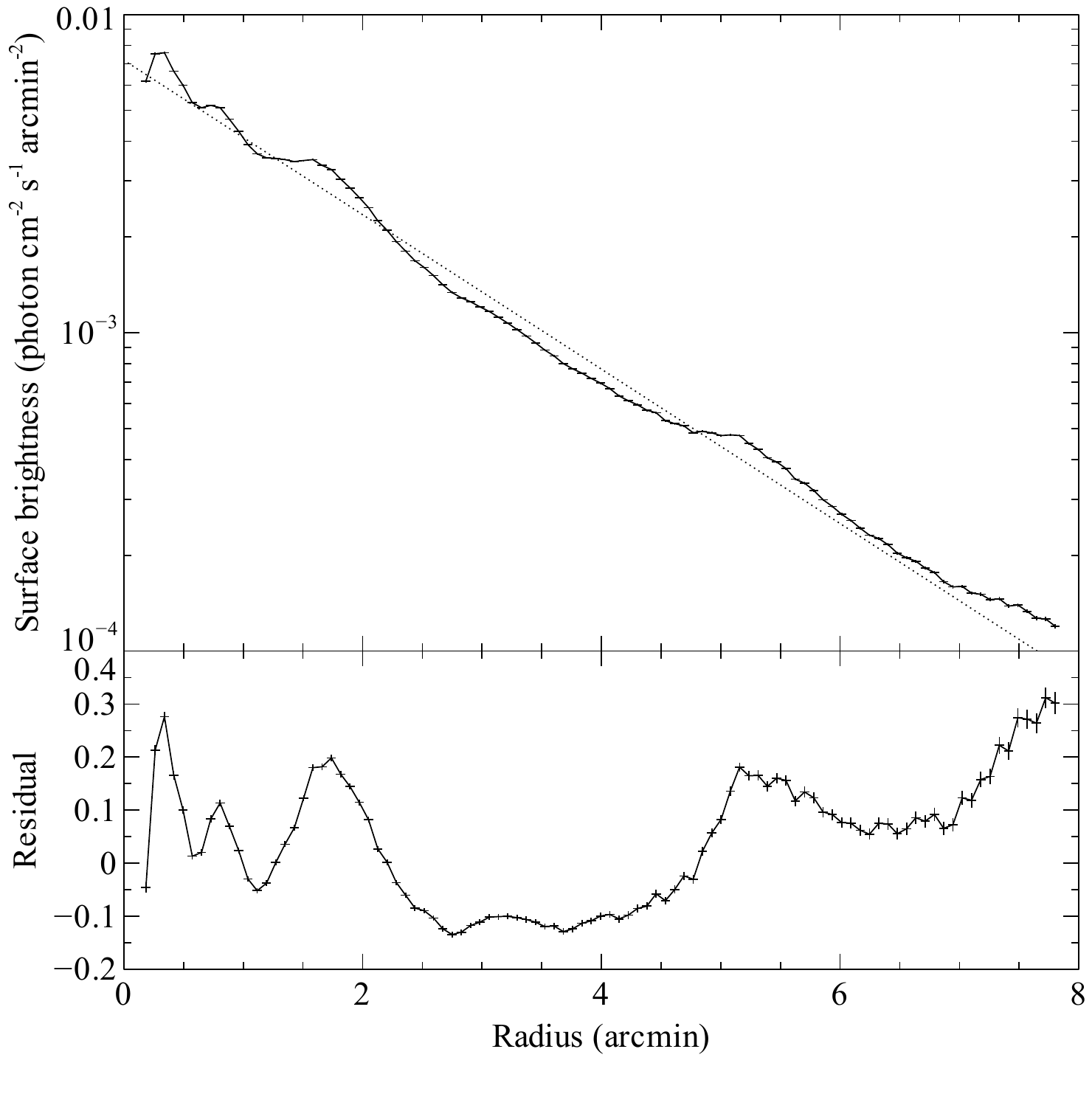}
  \caption{Surface brightness profile in the 0.5--7~keV band  to the
    West of the nucleus of NGC\,1275. }
\end{figure}

\begin{figure}
  \centering
  \includegraphics[width=0.45\textwidth,angle=0]{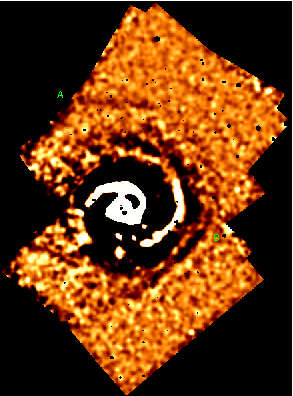}
  \caption{Unsharp masked image using larger regions than for Fig.~1. }
\end{figure}

\begin{figure}
  \centering
  \includegraphics[width=0.45\textwidth,angle=0]{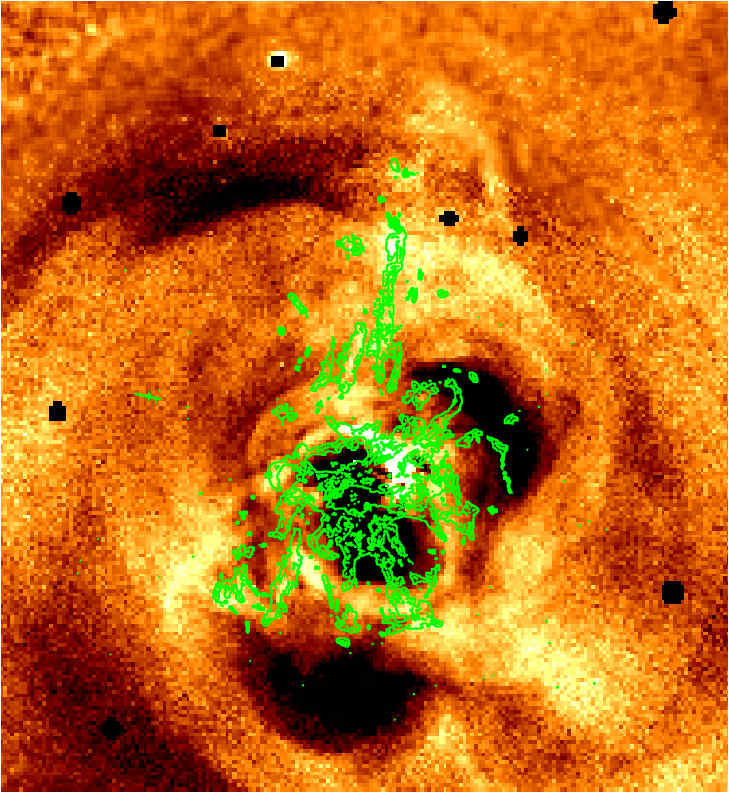}
  \caption{Inner part of the X-ray image with H$\alpha$ contours
    (from Conselice et al 2001) overlaid.  The image is 6.8 arcmin
    (150~kpc) from N to S.}
\end{figure}
\begin{figure*}
  \centering
  \includegraphics[width=0.9\textwidth,angle=0]{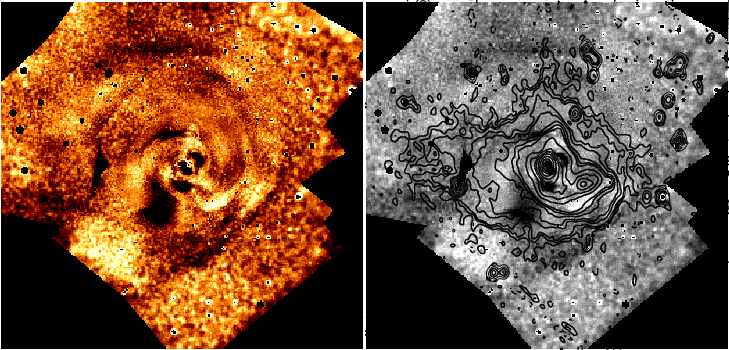}
  \caption{Left Chandra X-ray image residuals obtained after
    subtracting a smooth elliptical model.  Ellipses were fitted to
    surface brightness contours, and an elliptical model created by
    interpolating between these ellipses. Right: Radio contours (from
    Sijbring 1993) overlaid on the X-ray image. The radio data of 3C84
    were taken at 49~cm (608.5~MHz). Similar structures are seen in
    the 92~cm map. The radio contours stop abruptly at the edge of the
    outer ring of Xray emission to the SW. A spur of radio emission to
    the North points towards the trough. }
\end{figure*}

\begin{figure}
  \centering
  \includegraphics[width=0.45\textwidth,angle=0]{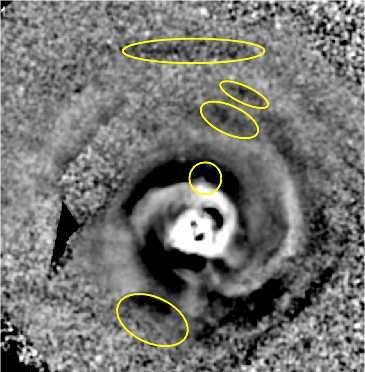}
   \caption{Structures identified in the composite image: from top to
     bottom --N
     trough, two older rising bubbles, top of N H$\alpha$ filament and
     S ``bay'' (old bubble?). }
\end{figure}
\begin{figure}
  \centering
   \includegraphics[width=0.45\textwidth,angle=0]{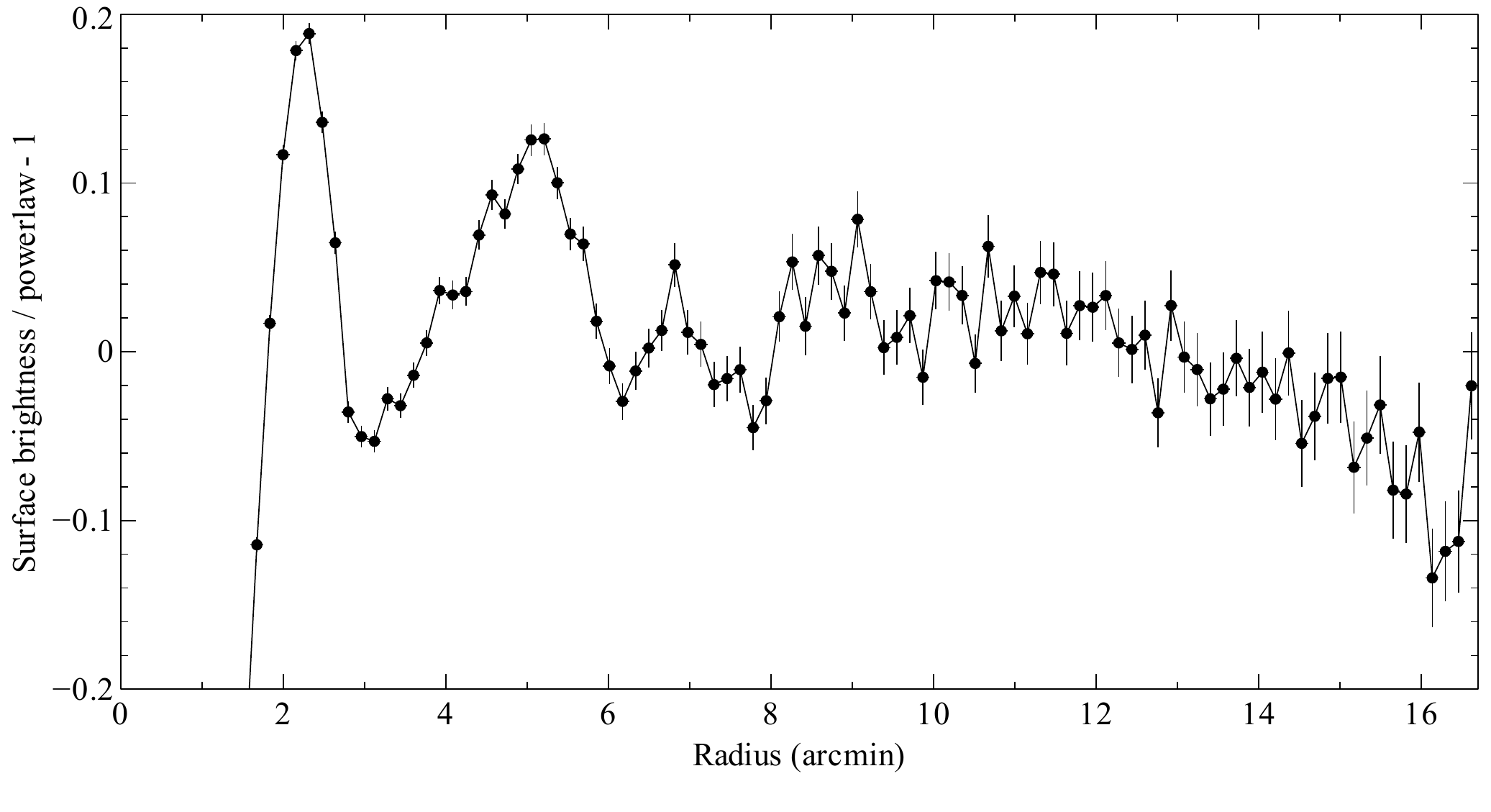}
  \caption{Surface brightness profile to the NNW. The dips at $\sim 6$
    and 7.7 arcmin are the two Northernmost bubbles. }
\end{figure}

We can start to explore a bubble origin by comparing the volume of the
trough with that of one of the observed bubbles near the centre. A
rough estimate of the volume of the trough can be made assuming that
it is a ellipsoid with its major axis, of length 2.5 arcmin, i.e. 55
kpc, in the plane of the sky and radius 0.7 arcmin, i.e. 15.5
kpc. This provides a 10 per cent dip in surface brightness, as
observed, if the integrated depth along the line of sight is
equivalent to the central value multiplied by the radius from the
nucleus. This volume is about $5.6 \times 10^4 \kpc^3$, or about 21.5
times that of the inner N bubble, which has a radius of 8.5~kpc.

The external pressure drops by a factor of 3 between the centre and 10
arcmin radius where the trough resides (Fig.~12), which is a pressure
of about $n_{\rm e} T_{\rm e}\sim 7 \times 10^5\K\pcm$.  Adiabatic
expansion then gives a factor of about 2 increase in volume, so we are
looking at the accumulation of about 11 bubbles similar to the present
inner ones (assuming the contents are preserved during the rise). The
energy content of the trough, $4PV=1.2\times 10^{60}\erg$. This is one and
two orders of magnitude smaller than the events that have occurred in
the Hydra A (Nulsen et al 2005) and MS\,0735.6+7421 (McNamara et al
2009) clusters, respectively.

The possibility arises that the trough represents the accumulation of
the last 500 Myr or so of bubbles (a similar structure is discussed
for A2204 by Sanders, Fabian \& Taylor 2009). Bubbles rise and are
trapped at some radius, in this case at about 220~kpc. Perhaps they
become neutrally buoyant there due to mixing with surrounding gas, or
the magnetic structure (possibly azimuthal there; Quataert et al 2008) traps
them. There also seems to be an overall structure at and just within
that radius to the W, possibly due to motion of the core relative to
the outer cluster gas (see e.g. Churazov et al 2003).

The two X-ray surface brightness dips to the SW of the trough
(Figs~9 and 10), which we identify as rising bubbles, have volumes of
approximately $10^4\kpc^3$ each, corresponding to about twice that of
the current inner bubbles.  

The bay to the South may result from the accumulation of Southward
rising bubbles in analogy to the Northern trough. It has a sharp,
curved Northern edge and the interior is hotter than the outer parts
(Fig.~11). Perhaps there has been some mixing and heating taking place
between the relativistic and thermal intracluster gases. It lies
much closer to the nucleus of NGC\'1275.

\begin{figure}
  \centering
 \includegraphics[width=0.45\textwidth,angle=0]{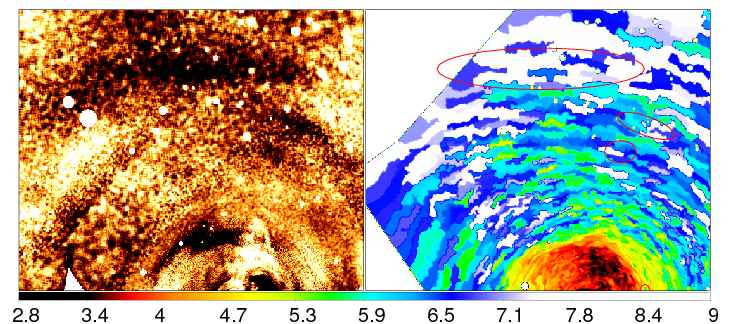}
  \includegraphics[width=0.45\textwidth,angle=0]{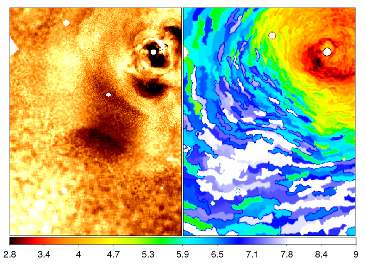}
 \caption{Top panel: Surface brightness and
    temperature maps to the N. Lower temperature gas is seen
    stretching S to the centre of the N trough. The two possible
    old bubbles coincide with higher temperature regions. The
    significance of this is not high since the temperatures have
    uncertainties of $\sim 0.5\keV)$ (???). Lower panel:  X-ray
    surface brightness with average at that
    radius subtracted (left) and temperature (right) to the SE,
    showing the ``bay'' which contains a ``tongue'' of hotter
    material. A sharp cold front occurs around the N side of the bay,
    but, being concave, is not a conventional cold front (Markevitch
    \& Vikhlinin 2008)   }
\end{figure}

The evolution of rising bubbles in cluster gas has been studied and simulated
by many authors (e.g. Diehl et al 2008; Liu et al 2008). Bubbles blown 
by a jet are not Rayleigh-Taylor unstable because the upper surface of
the bubble is not at rest relative to the hot gas above them. The
expansion of the bubbles means
that the hot gas continuously flows around them. The growth time of 
the Kelvin-Helmholtz
instability is comparable to the flow time. Whether they break up or
not depends on the amplitude and scale of velocity perturbations in
the hot gas. The stability of a large gas bubble rising through liquid
has been studied by Batchelor (1987). Rising air bubbles in water can
be surprisingly large. The scale size of disruptive
perturbations depends on surface tension (which in the
case of cluster bubbles means magnetic field), viscosity and
gravitational acceleration. Lyutikov
(2006) has proposed that magnetic draping can provide a surface
tension. Reynolds et al (2005) have discussed and simulated the role
of  viscosity in rising bubbles. Higher viscosity leads to stable,
long-lived, bubbles.

Diehl et al (2008) show that the bubbles seen at larger radii in several
clusters tend to be much
larger than those at smaller radii. Larger even than expected by
adiabatic expansion. In Hydra A, where 3 pairs of expanding bubbles
are identified (Wise et al 2007) there is a progressive steep increase of
bubbles radius with distance from the nucleus (see also Randall et al
2011 for a sequence of bubbles in an elliptical galaxy). 

Diehl et al (2008) discuss bubbles which are continuously fed energy
from the nucleus, which seems unlikely for Perseus. We note that if
the rising speed of a bubble depends on its size and if a merger
always occurs if two bubbles meet, then a sequence of bubbles of
slightly varying size will at larger radii become a set of fewer
larger bubbles. The radius at which this happens will depend on the
steepness of the bubble size-speed relation and the initial size
variation amplitude. It will happen both if larger ones are faster or
slower. This could well explain our observations of the outer Perseus
bubbles.

It is not clear what is happening to the S of the nucleus. The early
observations showed an inner and an outer bubble there. At larger
radii there is a sharp edge to part of an elliptical-shaped structure
which we term the Bay. This could represent the end point for bubbles
rising in that direction. The temperature there (Fig.~11) is
hotter. This could either be because there is hotter gas at that
radius, or it could be because relativistic plasma has displaced gas
and we are only seeing the projected outer hotter intracluster
gas. The pseudo-pressure maps (Fig.~13 and 14) do however peak in this
region so the first option is preferred. The energy in this cavity,
assuming it contains only thermal gas and a prolate-ellipsoidal shape, 
is $1.7\times 10^{60}\erg$, slightly more than for the Northern trough.

\begin{figure}
  \centering
  \includegraphics[width=0.45\textwidth,angle=0]{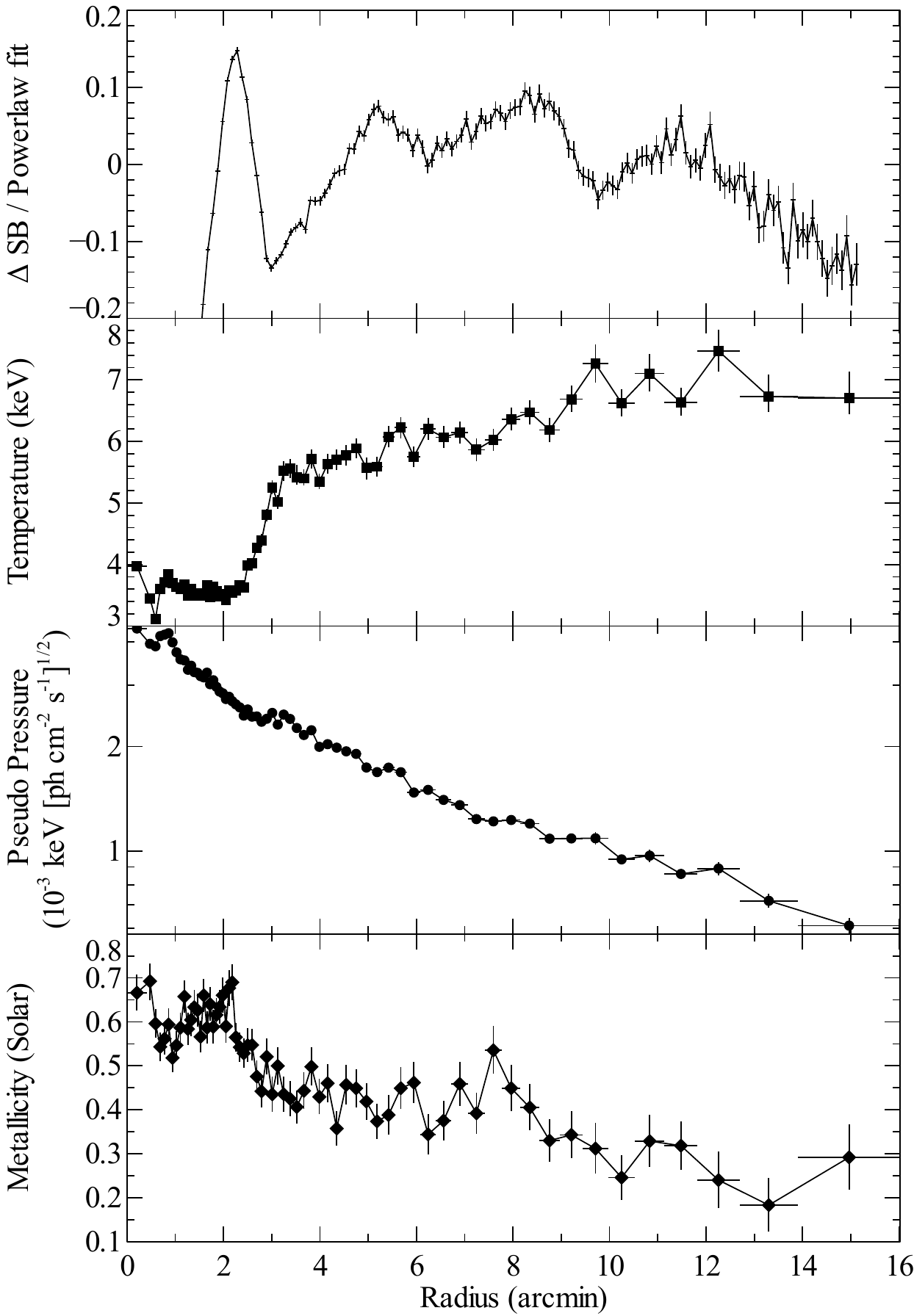}
  \includegraphics[width=0.45\textwidth,angle=0]{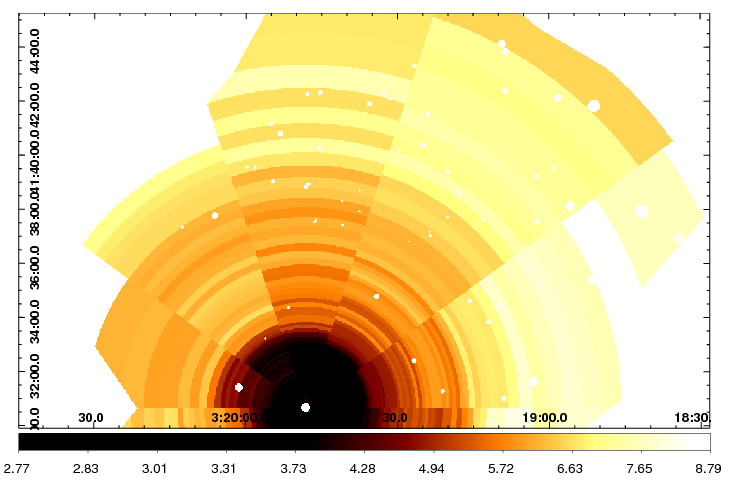}
  \caption{Profiles of differential surface brightness, temperature,
    psuedo-pressure and metallicity in the sector directly to the N
    (shown below). }
\end{figure}
\begin{figure*}
  \centering
  \includegraphics[width=0.98\textwidth,angle=0]{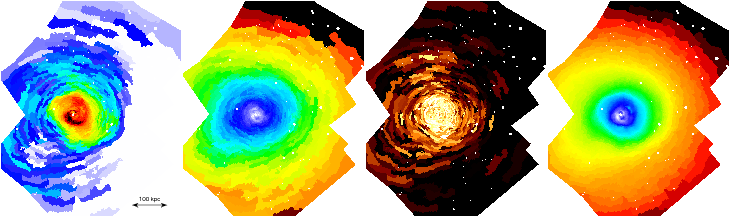}
  \caption{Distribution of (left to right) temperature,
    pseudo-pressure, metallicity and surface brightness. Each bin has
    about 22,500 counts.  }
\end{figure*}

\begin{figure*}
  \centering
  \includegraphics[width=0.9\textwidth,angle=0]{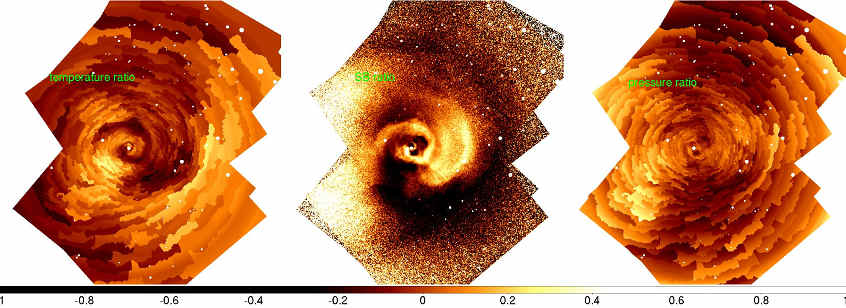}
  \caption{Temperature, surface brightness and pressure distributions.
  The means at each radius has been subtracted and the intensity is
  proportional to the fractional difference.}
\end{figure*}

\section{Temperatures, pressures and metals}

The new data have been contour-binned (Sanders 2007) into regions
each with 22,500 counts ($S/N=150$). The spectra within each bin have
been fitted separately resulting in the temperature and metallicity maps
shown in Fig.~13. The higher metallicity gas does stretch to the N
in the direction of the trough, but no more detailed correspondence
is evident. We also show a psuedo-pressure map (obtained from the
square root of the mean surface brightness in each region multiplied by the
temperature). 

A semicircular cold front is seen to the W. Across this front the
temperature jumps from 5.5-6~keV to 7 to 8~keV, and the metallicity
drops from 0.55-0.6 to 0.4-0.45. The pressure declines relatively
smoothly across the region (see also Figs.~16 and 17 in the
Appendix). The smooth pressure change while the temperature jumps is
characteristic of a cold front (see Section 6 for further discussion).

Finally, Fig.~14 shows the temperature and
pseudo-pressure maps obtained when the mean value at that radius (from
the nucleus of NGC\,1275) has been subtracted, to highlight small
differences.

\section{Radio correlations}

Minihaloes such as 3C84 are rare (Ferrari et al 2008) and may be due
to turbulence (Gitti et al 2004). There is a possible association
between minihaloes and gas sloshing (ZuHone \& Markevitch 2011) in
cluster cores (Mazzotta \& Giacintucci 2008). The swirl seen in the
temperature maps is indicative of gas sloshing, perhaps induced by an
off-axis merger or at least the close passage of a stripped group core
(Churazov et al 2003; Ascasibar \& Markevitch 2006; Roediger et al 2011). 

We have noted interesting correlations between structures in the radio
and X-ray maps (Fig.~8). While there might be small scale turbulence
in the hot gas, we doubt that there is considerable hydrodynamic
turbulence due to the straightness and ordered structures of the H$\alpha$
filaments. Indeed the 40--50~kpc long Northern filament shows little
evidence for turbulent flow. 

The large scale linearity of the bubble structures shown in Fig.~8,
apart from the Northern trough, is evidence against significant large
scale flows or rotation of the core gas. If we conservatively take a
minimum transverse motion of 50~kpc (the semi-ajor axis of the trough)
over the time to make the
structures, say $5\times 10^8\yr$, we obtain a maximum transverse velocity 
of $100\kmps$. 

\section{Large-scale stuctures and cold fronts}

The W edge, rim and semi-circular structure seen in Figs~1, 2, 5, and
6 appears to separate the inner AGN-dominated region from the outer
cluster. As outlined in Section 4, the temperature rises sharply to
the W of the structure, whereas the psuedo-pressure shows no abrupt
change (Figs~13 -- 16).  It therefore appears to be a cold-front (for
a review see Markevitch \& Vikhlinin 2007), and is probably related to
a past subcluster merger (Churazov et al 2003 and Section 5).

Mergers producing slopping/sloshing/bathtub/seich-modes in the core of
the Perseus cluster have been discussed for decades (see e.g. Allen et
al 1992), since Einstein Observatory imaging showed the overall
emission to be lop-sided (Branduardi-Raymont et al 1982). Such modes
can persist for a considerable time (Gyrs). Cold fronts can form in
the process (e.g. Birnoboim, Keshet \& Herquist 2010). Tangential
flows below the front and magnetic fields amplification are expected
(Keshet et al 2010). Once again we note the lack of evidence for
significant transverse flows in the relatively straight H$\alpha$
filament system or indeed in the NNW--SSW bubble axis, so any cold
front flows do not penetrate far. Radial velocity
measurements of the H$\alpha$ filaments show smooth laminar flow
(Hatch et al 2006; Salom\`e et al 2011).The radio mini-halo does end
abruptly at the front. The metallicity also undergoes a sharp change
(Fig.~13).

The Perseus cluster will be an excellent target for high resolution
X-ray spectroscopy, such as anticipated with ASTRO-H (Takahashi et al
2010). This will resolve the velocities flows along the line of sight.
Both the central region where the bubbles form (Heinz et al 2010) and
the W cold front will of great interest.

 \begin{figure}
  \centering
   \includegraphics[width=0.458\textwidth,angle=0]{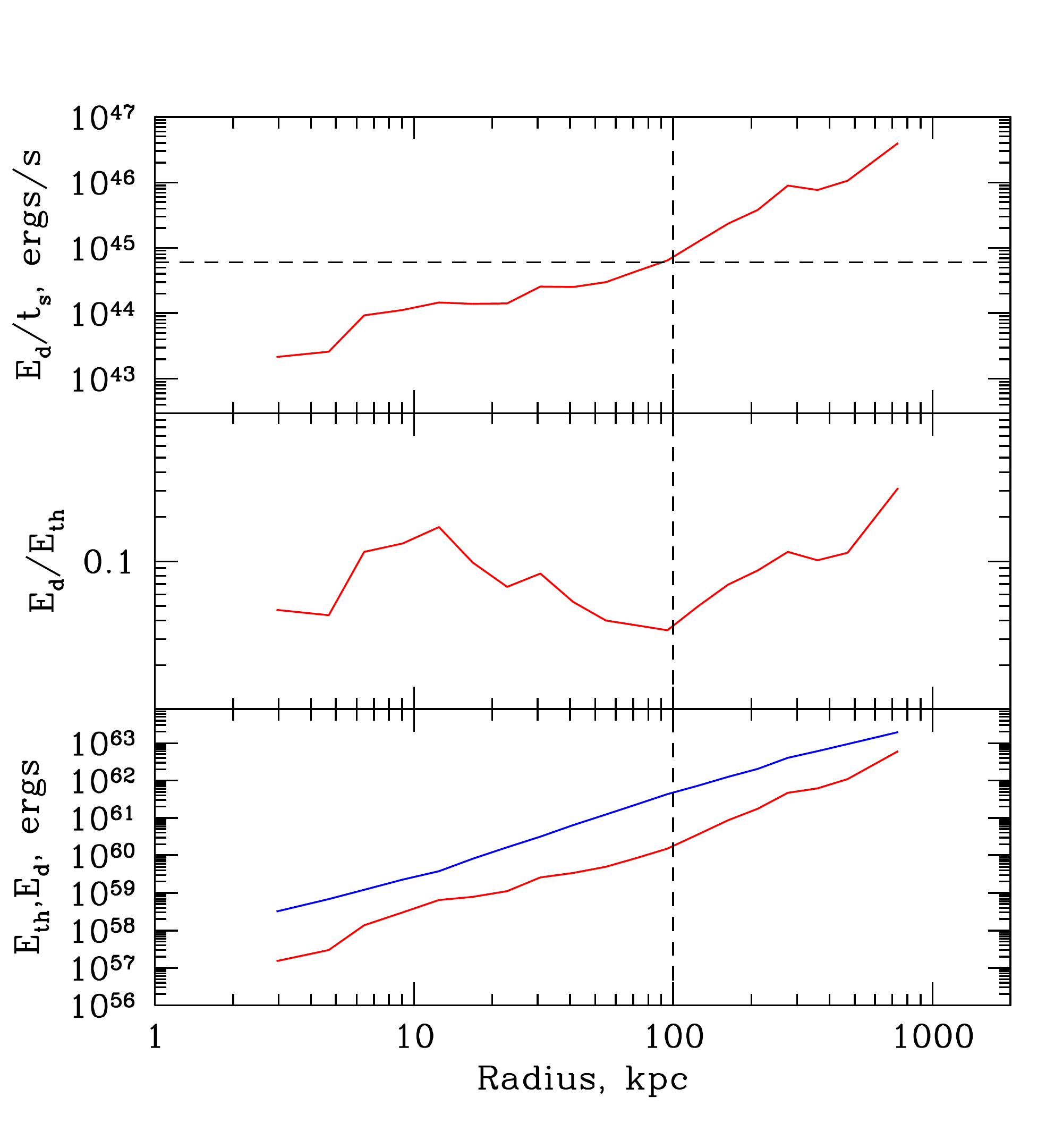}
  \caption{Energy and power associated with non-radial
  substructure. Bottom panel shows the total ICM thermal energy
   within given radius $E_{th}(<r)$  (see eqn.~1) and total energy
  associated with the observed substructure $E_{d}(<r)$ (eqn.~2). The
  ratio of these quantities is plotted in the middle panel. The top
  panel shows an estimate of power (eqn.~3) needed to support
  the observed non-radial substructure. The horizontal dashed line shows an
  estimate of the central AGN mechanical power, while the vertical
  line schematically divides the cluster into ``AGN-dominated'' and
  ``merger-dominated'' parts. }
\end{figure}

\subsection{The substructure profile}
In order to assess the energy associated with large scale structures
in the Perseus cluster, we have characterized the energy/power needed to
produce observed substructure (non-radial part) in the ICM
(e.g. Fig.~6).

In order to cover a large radial range we use XMM-Newton data. While
not providing a high resolution view on the Perseus core, they extend
to larger radii than the existing Chandra data\footnote{Analysis of
  the Chandra data gives similar results over the 10 -- 300~kpc range,
  but is noiser due to real structure at smaller radii and to
  photon noise at larger radii}. We first made a
deprojection analysis of the data in four wedges (90 deg each) and for
each wedge we calculated pressure in several radial shells spanning
the range from few to 700 kpc. The thermal pressure $nkT$ was
evaluated in each radial bin and root-mean-square variation of
pressure between wedges was calculated as $\delta P \displaystyle
\left \langle \left (P-\langle P(r)\rangle \right )^2 \right
\rangle^{1/2}$. The ratio of this quantity to the mean pressure at the
same radius $\displaystyle \frac{\delta P}{P}=\frac{\left \langle\left
      (P-\langle P(r)\rangle \right )^2 \right\rangle^{1/2}}{P}$ can
be used as a crude characteristic of the magnitude of the substructure
(non-radial part) in the cluster. We then calculated the total thermal
energy $E_{th}(<r)$ of the ICM within given radius
\begin{eqnarray}
E_{th}(<r)=\int_{0}^{r} \frac{3}{2}nkT 4\pi r^2dr,
\label{eq:eth}
\end{eqnarray}
where $n$ and $T$ are the gas density and temperature at a given
radius and estimated the energy associated with the substructure $E_{d}(<r)$ as
\begin{eqnarray}
E_{d}(<r)=\int_{0}^{r} \frac{3}{2}nkT \frac{\delta P}{P} 4\pi r^2dr.
\label{eq:ed}
\end{eqnarray}
The values of $E_{th}(<r)$ and $E_{d}(<r)$ and their ratio are plotted
in the two bottom panels in Fig.~15. In the top panel the energy
associated with substructure $E_{d}(<r)$ is divided by the sound
crossing time of the region $t_s=r/c_s$, where $\displaystyle
c_s=\sqrt{\gamma \frac{kT}{\mu m_p}}$. The quantity
\begin{eqnarray}
L_d=\frac{E_{d}(<r)}{t_s} 
\label{eq:ld}
\end{eqnarray}
can be regarded as an estimate of the power needed to maintain the
observed non-radial substructure.

Clearly these are order of magnitude estimates. By construction they
are sensitive only to non-radial part of the substructure and only to
the low angular modes. Even more uncertain is the estimate of power
(top panel in Fig.~15) which uses the radial sound crossing time as an
estimate of the time needed to dissolve the
substructure. Nevertheless, taken at face value this figure suggests
that the cluster can be broadly divided into two zones at about 100
kpc. (The bright rim to the W of the nucleus is at a radius of $\sim
110\kpc$, see Figs.~2 and 6).The estimates of the mechanical power of
the central AGN in NGC1275 gives values of order $10^{45}\ergps$ over
the period of $10^8\yr$ (Fabian et al 2006). If the same estimate is
applicable to the mean power during the last Gyr, then the central AGN
is not capable of producing the asymmetry outside the central
$\sim100\kpc$ region. Instead the most promising explanation for the
substructure beyond 100 kpc is the merger of the Perseus cluster with
smaller galaxies/group along the chain of bright galaxies to the West
of the core. At yet larger radii beyond 1~Mpc and out to the virial
radius, the intracluster gas appears to become become intrinsically
clumpy (Simionescu et al 2011).

\section{Summary}

Our wider image has revealed several structures consisting of dips in
surface brightness along the NNW--SSE axis which are plausibly outer
rising bubbles. The two new ones are not much larger in
pressure-corrected volume than the inner bubbles whereas the large
trough to the North is an order of magnitude larger.  Bubbles must be
long-lived and we suggest that they may grow in size by merger with
other, slower, rising bubbles.

A roughly semi-circular cold front is see to the W of the nucleus of
NGC\,1275. The metallicity of the gas drops abruptly across the front.  

We suspect that the trough and a bay to the South are where rising
bubbles have accumulated. The S bay is distinctly hotter than its
surroundings so the relativistic plasma may have mixed with the
intracluster gas there, unlike the situation of the N or the inner
bubbles. Both the N trough and the S bay lie along a continuation of
the W cold front.  The other structures seen at radii or 100 - 150~kpc
are likely associated with a subcluster merger (Churazov et al
2003). The merger also accounts for the E--W asymmetry in surface
brightness across the image (Figs.~2, 4, 14).

The energy in substructure peaks at small radii, where it is dominated
by activity induced by the AGN, and at larger radii, where it is
dominated by the merger. 

\section*{Acknowledgements}
We acknowledge financial support from the Royal Society (ACF). JHK
thanks the Cambridge Trusts and the Natural Sciences and Engineering
Research Council of Canada (NSERC). GBT acknowledges support for this
provided by the National Aeronautics and Space Administration through
Chandra Award Numbers GO0-11139X and GO0-11138B issued by the Chandra
X-ray Observatory Center, which is operated by the Smithsonian
Astrophysical Observatory for and on behalf of the National
Aeronautics Space Administration under contract NAS8-03060. CSR
acknowledges support from the Chandra Guest Observer Program under
grant GO011138A. SWA was supported in part by the U.S. Department of
Energy under contract number DE-AC02-76SF00515.


\section{Appendix}

Temperature and pressure profiles about the nucleus in 18~deg steps
anticlockwise from W.

\begin{figure*}
  \centering
  \includegraphics[width=0.8\textwidth,angle=0]{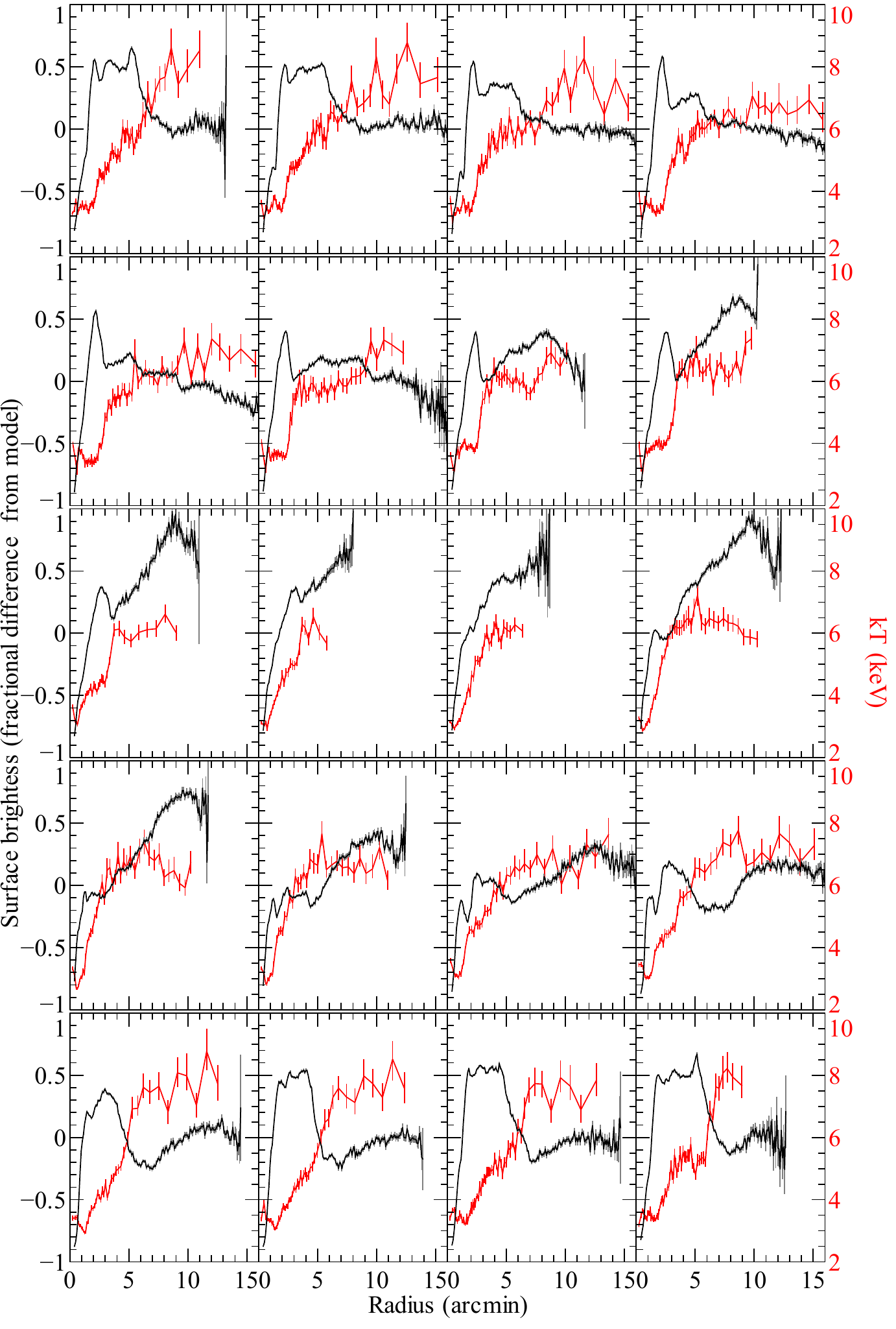}
  \caption{Temperature profiles (in red) with differential surface brightness
    (black; subtracted from a smooth model) in 20x18 deg sectors
    arranged from W through N. }
\end{figure*}

\begin{figure*}
  \centering
  \includegraphics[width=0.8\textwidth,angle=0]{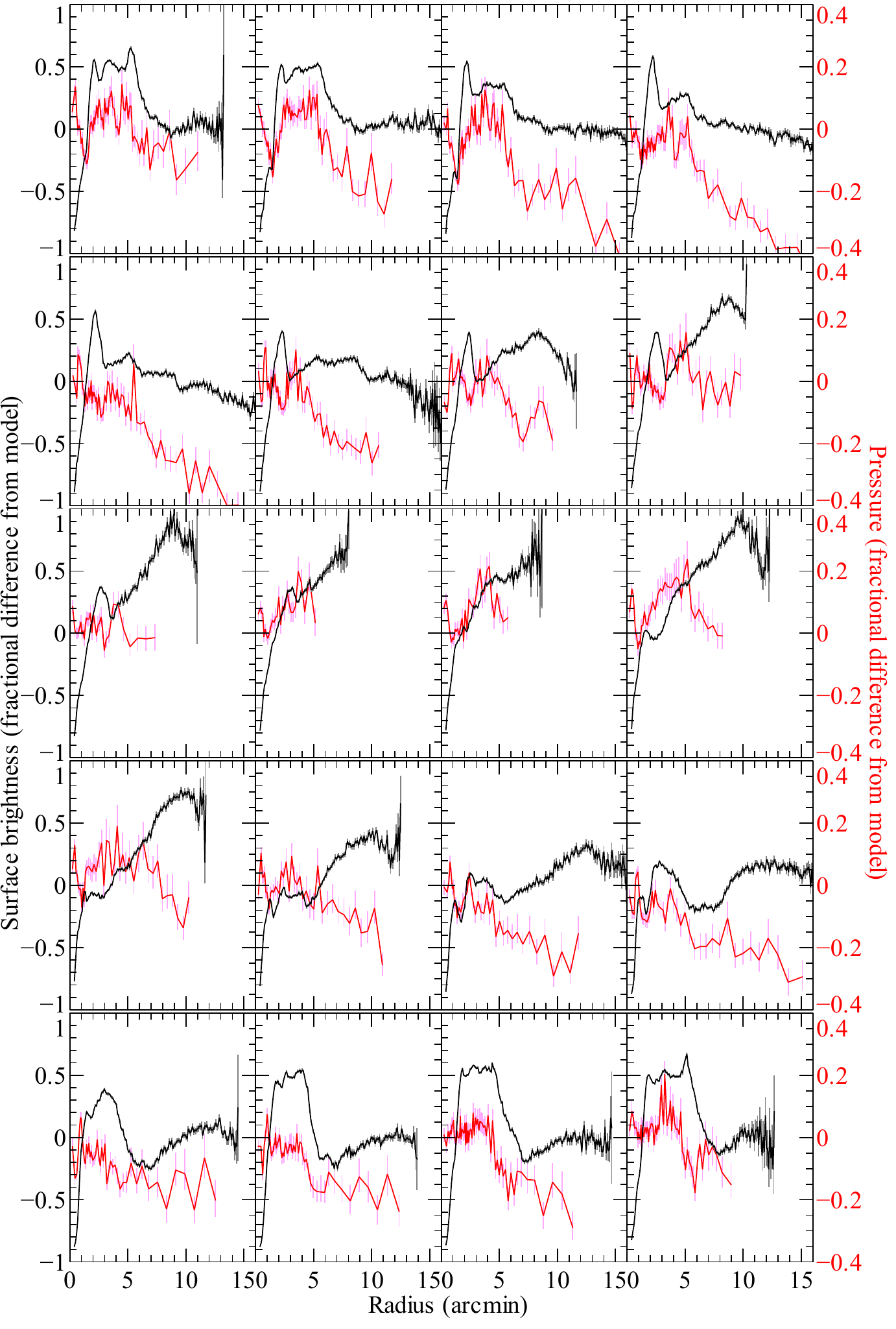}
  \caption{Similar to Fig.~4, but for Pressure (the model is a
    $\beta$-profile with a core, which explains why these profiles are
    somewhat different near the centre to the other profiles
    e.g. Fig~5, which use a power-law model at all radii). }
\end{figure*}

\end{document}